# Transfer learning for cancer diagnosis in histopathological images

Sandhya Aneja[1], Nagender Aneja[2], Pg Emeroylariffion Abas[1], Abdul Ghani Naim[2]
[1]Faculty of Integrated Technologies, Universiti Brunei Darussalam, Bandar Seri Begawan, Brunei Darussalam
[2]School of Digital Science, Universiti Brunei Darussalam, Bandar Seri Begawan, Brunei Darussalam

**ABSTRACT**

Transfer learning allows us to exploit knowledge gained from one task to assist in solving another but relevant task. In modern computer vision research, the question is which architecture performs better for a given dataset. In this paper, we compare the performance of 14 pre-trained ImageNet models on the histopathologic cancer detection dataset, where each model has been configured as naive model, feature extractor model, or fine-tuned model. Densenet161 has been shown to have high precision whilst Resnet101 has a high recall. A high precision model is suitable to be used when follow-up examination cost is high, whilst low precision but a high recall/sensitivity model can be used when the cost of follow-up examination is low. Results also show that transfer learning helps to converge a model faster.

*Corresponding Author:*

Nagender Aneja
School of Digital Science, Universiti Brunei Darussalam
Bandar Seri Begawan, Brunei Darussalam
Email: nagender.aneja@ubd.edu.bn

## 1. INTRODUCTION

Transfer learning focuses on transferring latent knowledge from a source domain to a target domain, to solve the issue of insufficient data on the target domain. In the case of deep transfer learning, knowledge comprises architecture and trained parameters, wherein a nonlinear function is learned for the task in the target domain. Transfer learning has been used in computer vision, natural language processing, speech recognition, and deep reinforcement learning and has been successfully deployed in industries, including health care, autonomous driving, robotics, spam filtering, search, and recommendation. Applications of transfer learning in convolutional neural network (CNN) provide the benefits of starting the training process with a better initial model, and thus, faster learning and better performance. Transfer learning also is a useful approach for data and resource-constrained applications.

Although transfer learning is advantageous in transferring features from a source domain to a target domain, it may negatively impact the performance in the target domain, especially if the source domain is not related to the target domain. This is also known as negative transfer [1]. Recently, a number of CNN architectures have been proposed, with varying width i.e., number of feature maps, and depth i.e., number of layers. The architectures also vary in other parameters such as connections between hidden layers, normalization, and non-linear function. Most of the architectures have been shown to perform well on the ImageNet dataset. As such, for a specific application, there is a need to study which of the existing network architectures generalize well on that particular application.

Transfer learning is an extension of supervised learning such that the training set is from a source domain while the test set is from a target domain [2]. In traditional supervised learning, it is assumed that the test set comes from the same joint probability distribution of features and labels, and the posterior probability $P(X \cap Y) = P(Y|X).P(X)$ can be learned. However, transfer learning relaxes the condition that the test is from the same distribution and may not have labels or the same labels. A number of studies have been published on machine learning on cancer data, as it is the second-leading cause of death in the world. Mahmood *et al.* [3] surveyed neural network techniques for the classification of different cancers, where it has been observed that on many occasions, practitioners' opinion on the detection of a disease may vary according to clinical facts. A computer-based analysis would certainly help practitioners, by providing a second expert-opinion since neural network mimics the brain process. A carefully selected training data would be needed to accurately predict the disease.

Kornblith *et al.* [4] presented performance of 16 networks on non-health datasets. The authors found a correlation between ImageNet Accuracy and transfer accuracy when used as a fixed feature extractor or fine-tuned, however, the correlation has been found to be sensitive, in the case of a fixed feature extractor. The authors also discovered that in the case of Stanford Cars and fine-grained visual classification (FGVC) Aircraft dataset, pretraining using ImageNet provides minimal benefits. Tajbakhsh *et al.* [5] presented work on transfer learning in classification, detection, and segmentation between natural and medical images from radiology, cardiology, and gastroenterology. The authors demonstrated that pretraining with fine-tuning outperforms CNN that has been trained from scratch. Layer-wise tuning has also been shown to offer the best performance.

Guo *et al.* [6] presented an algorithm to find an optimal strategy for transfer learning for each instance, whereby the authors have used a policy network to decide whether to pass the image through fine-tuned layers or pre-trained layers. Opbroek *et al.* [7] have shown that support vector machine (SVM) based classifiers can outperform and minimize classification error by 60%, with a small amount of training data by using transfer learning. Raghu *et al.* [8] explored properties of transfer learning for medical imaging datasets, pertaining to Retina and CheXpert (chest x-ray images) data, and have shown that transfer learning offers little benefit to performance, with lightweight models giving comparable performance to the more complicated ImageNet architectures. However, the authors have only used two architectures: ResNet50 and Inception-v3, in their analysis. Cheplygina *et al.* [9] discussed around 140 papers on semi-supervised, multi-instance, and transfer learning, and have found transfer learning as the most popular. The usefulness of transfer learning has been demonstrated, but also mentioned the need for future research to examine connections between learning scenarios and generalization of results.

Rustam *et al.* [10] applied regression logistics and random forest for pancreatic cancer detection and have found random forest to perform better. Al-Khowarizmi and Suherman [11] applied a simple evolving connectionist system (SECoS) on a balanced cancer dataset, whilst Dabeer *et al.* [12] applied CNN for breast cancer detection using histopathology images from a biopsy. Pratiwi *et al.* [13] ensembled three architectures: inception V3, inception ResNet V2, and DenseNet 201, with the ensembled model showing sensitivity of 90%, specificity of 97%, precision of 82%, and recall of 85%. Thus, a comprehensive study of major architectures will be helpful for researchers to select the architecture for other datasets also.

## 2. RESEARCH METHOD

Transfer learning is an optimization technique to apply learning of one task to another task. This is important since deep learning needs lot of data to train and collecting data for each task is a challenging problem. Transfer learning can be applied in various ways. The section below provides a few definitions to formally define transfer learning, as well as presents various transfer learning methods.

− Definition 2.1 (Feature Space). Feature space, *X*, is set of all possible images of a particular size where each pixel value ranges 0-255. Training set is assumed to be subset of feature space, $X = \{x^i, x^i \in X, 1 \leq i \leq n\}$.
− Definition 2.2 (Label Space). Label space, *Y*, includes all possible labels for samples, and $Y = \{y^i, y^i \in Y, 1 \leq i \leq n\}$. In this paper, *Y = {0, 1}*, where 1 indicates presence of cancer tissue and 0 otherwise.
− Definition 2.3 (Prediction Function). Our objective is to learn prediction function $\hat{y} = f(x) \, or \, P(\hat{y}|x)$.
− Definition 2.4 (Domain). Domain, *D*, is set of feature space and probability distribution of training set $D = \{X, P(X)\}$.
− Definition 2.5 (Task). Task, *T*, is set of label space to predict and prediction function $T = \{Y, P(Y|X)\}$.
− Definition 2.6 (Transfer Learning). Given source domain $D_S = \{X_S, P(X_S)\}$ and source learning task $T_S = \{Y_S, P(Y_S|X_S)\}$, and target domain $D_T = \{X_T, P(X_T)\}$ and target learning task $T_T = \{Y_T, P(Y_T|X_T)\}$, transfer learning deals with learning prediction function of target $P(Y_T|X_T)$ using knowledge of source domain $D_S$ and source task $T_S$ such that $D_S \neq D_T, T_S \neq T_T$, or no or limited labels in target domain [14].

## 2.1. Classification of deep transfer learning approaches

Transfer learning can help to provide a better initial model that can be used to faster training fro a different task. Neural information processing systems (NIPS) 1995 workshop learning to learn: knowledge consolidation and transfer in inductive systems provided initial motivations for this. Transfer learning approaches can be classified based on Domain and Task; Feature space and label space; Source and Target label space; and knowledge transfer process as described below:

### 2.1.1. Domain and task

Transfer of knowledge is possible in a wide variety of domains. Transfer Learning aims to improve learning of target task on target domain using knowledge from source task on source domain [15]. Transfer Learning is sometimes referred as domain adaption. We can classify transfer learning techniques, based on domain and task:

a. Same Domain and Different Task. Since the task is defined by $T = \{Y, P(Y|X)\}$, either $Y$ is different or $P(Y|X)$ is different for source and target domain (or both are different).
- $Y_S \neq Y_T$, e.g., the source data set is CIFAR-10 whilst the target data set is CIFAR-100,
- $P(Y_S|X_S) \neq P(Y_T|X_T)$ i.e., the same image may be labeled differently in the source and the target domain, and it happens when the class balance is different (prior $P(Y)$ is different in source and target.)

b. Different domain and same task. Since the domain is defined by $D = \{X, P(X)\}$, either $X$ is different or $P(X)$ is different for source and target domain (or both are different)
- $X_S \neq X_T$, different feature space, e.g., source data set is gray-scale images whilst target data set is color images or documents in different languages, $P(X_S) \neq P(Y_T)$, different distributions, while feature space may be same, e.g., the source domain consists of hand-drawn images while the target domain consists of photographs or documents in the same language but different topics. This is the most common scenario, commonly referred to as domain adaption, e.g. An autonomous model is trained for day driving but the trained model applied in all other scenarios.

### 2.1.2. Feature space and label space

Zhuang *et al.* [16] proposed a classification of transfer learning as homogeneous and heterogeneous transfer learning, based on feature space $X$ and label space $Y$. Homogeneous transfer learning means $X_S=X_T$ and $Y_S=Y_T$. Heterogeneous transfer learning means $X_S \neq X_T$ or/and $Y_S \neq Y_T$.

### 2.1.3. Source and target label space

Pan and Yang [14] classified transfer learning as Transductive, Inductive, and Unsupervised transfer learning, based on the label space $Y$. Transductive transfer learning means $Y_S$ exists but $Y_T$ does not exist. Inductive transfer learning means both $Y_S$ and $Y_T$ exist. \Unsupervised transfer learning means both $Y_S$ and $Y_T$ do not exist.

### 2.1.4. Knowledge transfer process

Tan *et al.* [17] classified deep transfer learning in four categories based on knowledge transfer process, namely, instances based, mapping based, network-based, and adversarial based. On the other hand, Li *et al.* [18] classified deep transfer learning into three approaches: instance transfer, model transfer, and feature representation transfer.

a. Instances-based or instance transfer: selects supplementary data from the source domain, whereby the selected source instances are similar to instances in the target domain. This is different from generative data augmentation, which generates synthetic data using generative adversarial network (GAN). Instance based transfer learning avoids negative transfer by selecting similar instances.
b. Mapping-based: both source and domain instances are mapped to a third domain space, whereby both domains are similar to each other. A union deep neural network is trained on modified source and target domain.
c. Network-based or model/feature representation transfer: this is the most common approach, and it focuses on network architecture and/or trained parameters. Source network architecture may be used as a feature extractor wherein convolution layers are frozen to extract the features, or may be used to fine-tune the target model, whereby the target model is initialized with trained parameters from the source domain, and then few/all of the layers are subsequently trained for the target domain. In other cases, only the source network architecture is used with random weights for the target domain. This approach can further be classified in four cases:
- Target data set is large but different from the source: in this case, the entire model is trained for the target domain using sequential training for the source domain or joint training for both source and target domain
- Target data set is large and similar to the source: in this case, some layers are trained while other layers are frozen
- Target data set is small but different from source: in this case, some layers are trained while leaving other layers frozen and/or the use of domain adaption to find common feature space, which may also be determined using GANs [19], and
- Target data set is small and similar to the source: in this case, convolution base is frozen (no adaption to target).

d. Adversarial-based: In this case, transferable representations are determined using GAN. These transformations are applicable to both source and target domains. The features from both domains are extracted using front layers and sent to adversarial layers. The adversarial layer is trained to discriminate origin of features. Thus, the network is trained to learn general features. The representations are called transferable when the features are discriminative for the source domain, and indiscriminate with the shift between source and target domains [20].

In this paper, network-based or model/feature representation transfer for cancer detection has been analyzed. In particular, PatchCamelyon (PCam) benchmark data set for histopathologic cancer detection competition organized by Kaggle [21], has been used. The objective of the competition was to create an algorithm to identify metastatic cancer in small image patches taken from larger digital pathology scans.

The dataset [21] is a modified version of the PatchCamelyon (PCam) benchmark dataset [22], [23], with duplicate images removed. It has a total of 220,025 images, with 130,908 (60%) benign images and 89,117 (40%) images, where the center 32×32 px region of a patch contains at least one pixel of tumor tissue. Thus, the dataset is not balanced. Eight sample images from the dataset, containing tissue images of healthy and patient with tumor, are shown in Figure 1. Digitization of microscopic glass slide images has encouraged deep learning researchers for pathology diagnosis, where manual results differ considerably. In particular, trained deep convolutional neural network has been shown to perform better than manual cancer detection by different Pathologists. In this paper, performance of transfer learning from 14 CNN architectures has been presented. research methodology comprises three model training approaches for each architecture. The three training approaches are naive, feature extraction, and fine-tuning. These approaches are based on neural network architecture that comprises of convolutional layers and fully-connected layer(s).

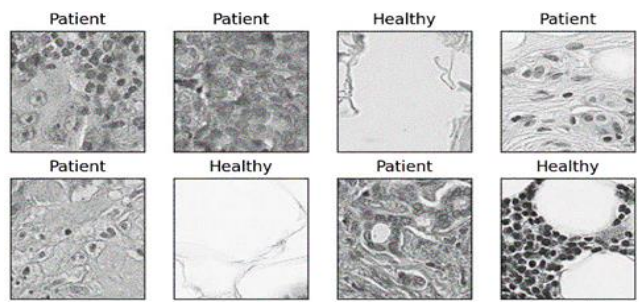

Figure 1. Sample images of healthy and patient with tumor, with center 32×32 px containing tumor tissue

In the Naive training experiment, neither the convolutional layers nor the fully connected layer(s) are trained, with only the pre-trained (trained on ImageNet dataset) layers used for testing. On the other hand, in the feature extractor experiment, the fully connected layer is trained for five epochs, whilst freezing the convolutional layers. Finally, in the fine-tune experiment, all layers from the convolutional layers and fully connected layer are trained. The fine-tune experiment is conducted for five epochs, only after conducting the second feature extractor experiment for five epochs. It has been found training for five epochs are sufficient for the model to converge, for both the convolutional and fully connected layers. At every epoch, we saved the model based on error rate i.e., after five epochs, we get the final model with the lowest error rate. The final saved model with the lowest error rate has been used to evaluate and compare different architectures.

## 3. RESULTS AND DISCUSSION

14 modern deep convolutional neural networks: Alexnet, Vgg16 bn, Vgg19 bn, Resnet18, Resnet34, Resnet50, Resnet101, Resnet152, Squeezenet1 0, Squeezenet1 1, Densenet121, Densenet169, Densenet201, and Densenet161, pretrained on ImageNet have been analyzed. We have considered batch size 32, input size 196×196, 80% training data, and 20% validation data, and learning rate of 0.01. Although the recommended input size is 224×224 for pre-trained models, we have reduced the input size slightly to 196×196 so that all models can fit onto our Nvidia 1080 GPU memory. Increasing input size is expected to increase Area Under the receiver operating characteristic curve (AUC) Score. Three experiments with 14 models, thus, a total of 42 sub-experiments, have been performed for this study.

### 3.1. Naive model

In the naive model, we have used weights from the pre-trained network. Since the model was not trained for the target task on the target domain, the model didn't generalize well. This will happen in particular when the target domain is different from source somain as is the case in this research. Results as in Figure 2 indicate that most models do not generalize well.

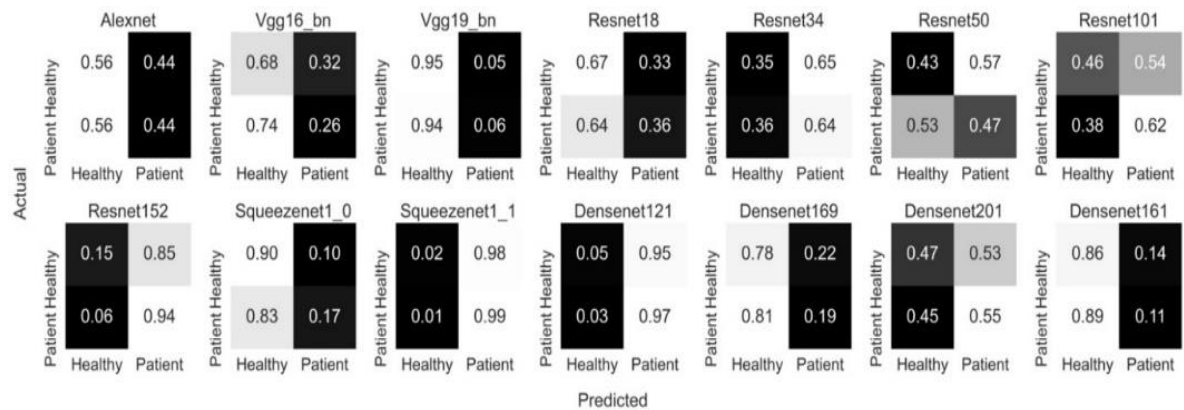

Figure 2. Naive model: confusion matrix (normalized)

### 3.2. Feature extractor model

In the feature extractor model, convolution layers are frozen, and the fully connected layer is trained for five epochs. Figure 3 displays the normalized confusion matrix after five epochs using the model that gives the lowest error rate during five epochs for every architecture. Figure 3 shows confusion matrix and Table 1 shows the performance of every architecture using the feature extractor model, in terms of AUC Score, F1 Score, Precision, and Recall. It can be seen that Densenet161 represents a high precision model, with precision value of 0.9710, while Resnet101 represents a high recall model, with recall value of 0.9569.

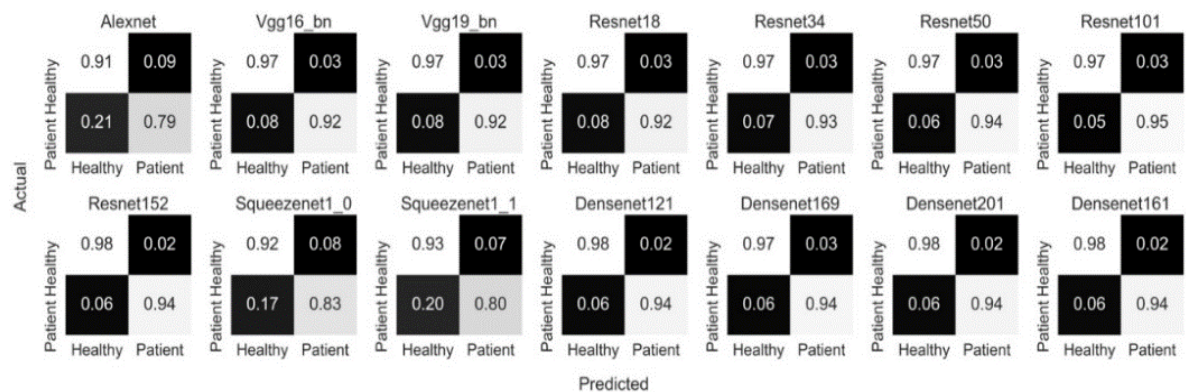

Figure 3. Feature extractor model: confusion matrix (normalized)

Table 1. Feature extractor model: AUC score, F1, precision, and recall for the 14 model

|  | AUC Score | F1 | Precision | Recall |
|---|---|---|---|---|
| Alexnet | 0.93872 | 0.82750 | 0.87121 | 0.78796 |
| Vgg16 bn | 0.98606 | 0.93751 | 0.95044 | 0.92492 |
| Vgg19 bn | 0.98710 | 0.93987 | 0.95831 | 0.92212 |
| Resnet18 | 0.98640 | 0.93700 | 0.95482 | 0.91983 |
| Resnet34 | 0.98823 | 0.94438 | 0.95614 | 0.93292 |
| Resnet50 | 0.99046 | 0.94923 | 0.95990 | 0.93879 |
| Resnet101 | 0.99131 | 0.95106 | 0.95685 | 0.94534 |
| Resnet152 | 0.99131 | 0.95256 | 0.96525 | 0.94019 |
| Squeezenet1 0 | 0.94847 | 0.85222 | 0.87606 | 0.82964 |
| Squeezenet1 1 | 0.94659 | 0.84265 | 0.89129 | 0.79904 |
| Densenet121 | 0.99065 | 0.95115 | 0.96355 | 0.93907 |
| Densenet169 | 0.99151 | 0.95181 | 0.96173 | 0.94209 |
| Densenet201 | 0.99196 | 0.95468 | 0.96566 | 0.94394 |
| Densenet161 | 0.99240 | 0.95656 | 0.97101 | 0.94254 |

### 3.2. Fine tune model

In the fine tune model, we run five epochs to train the full network in addition to training a fully connected layer, with the fine-tuning experiment done only after feature extraction. Figure 4 displays normalized confusion matrix after five epochs using the models that give the lowest error rate. For this experiment, surprisingly, Squeezenet1 1 gave the worst performance, predicting most subjects as patients with cancer tissue as shown in the confusion matrix. Squeezenet1 1 has smaller number of parameters than the number of parameters in Squeezenet1 0 and other models, due to model's compression technique, as given in Iandola et al. [24], [25]. This trade-off has resulted in reduced performance of the model, as observed in the Figure 4. Table 2 shows that Densenet161 has a high precision value of 0.9703, whilst Densenet201 has a high recall value of 0.9458.

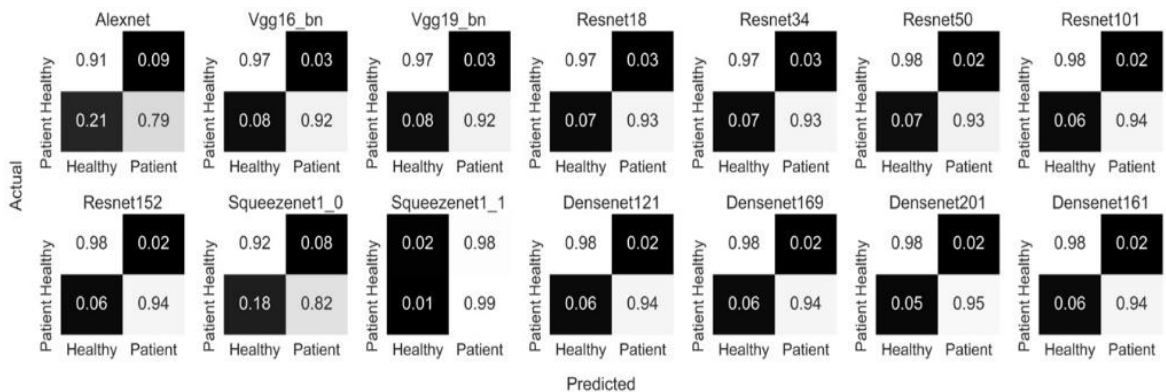

Figure 4. Fine tune model: confusion matrix (normalized)

Table 2. Fine tune model: AUC score, F1, precision, and recall for the 14 models

|  | AUC Score | F1 | Precision | Recall |
|---|---|---|---|---|
| Alexnet | 0.93362 | 0.82175 | 0.85449 | 0.79143 |
| Vgg16 bn | 0.98600 | 0.93684 | 0.95306 | 0.92117 |
| Vgg19 bn | 0.98707 | 0.93853 | 0.96015 | 0.91787 |
| Resnet18 | 0.98630 | 0.93874 | 0.95239 | 0.92548 |
| Resnet34 | 0.98844 | 0.94409 | 0.96111 | 0.92766 |
| Resnet50 | 0.99097 | 0.94946 | 0.96852 | 0.93113 |
| Resnet101 | 0.99149 | 0.95133 | 0.96646 | 0.93667 |
| Resnet152 | 0.99080 | 0.95192 | 0.96612 | 0.93812 |
| Squeezenet1 0 | 0.94676 | 0.85020 | 0.87813 | 0.82399 |
| Squeezenet1 1 | 0.52655 | 0.57968 | 0.40978 | 0.99027 |
| Densenet121 | 0.99092 | 0.95146 | 0.96277 | 0.94042 |
| Densenet169 | 0.99186 | 0.95321 | 0.96653 | 0.94025 |
| Densenet201 | 0.99235 | 0.95490 | 0.96418 | 0.94579 |
| Densenet161 | 0.99225 | 0.95520 | 0.97028 | 0.94058 |

## 4. CONCLUSION

As humans learn new things by applying knowledge from previous learnings, transfer learning can help the progress of artificial general intelligence further. This research has shown that Densenet161 outperforms other architectures, with an AUC score of 0.9924 and F1 Sore 0.95 in the feature extractor model. Densenet161 has been shown to have high precision whilst Resnet101 is a high recall model. No significant performance improvement has been observed through fine-tuning, as the model has been first trained using feature extraction, such that there is limited scope for further improvement. A high precision model may be used where follow-up examination cost is high, whilst low precision but a high recall/sensitivity model can be used when the cost of follow-up examination is low. Future work may include applying pre-trained models on other categories of health datasets.

## BIOGRAPHIES OF AUTHORS

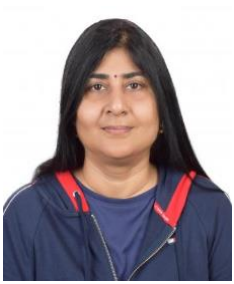 **Dr Sandhya Aneja** is working as Assistant Professor of Information and Communication System Engineering at the Faculty of Integrated Technologies, Universiti Brunei Darussalam. Her primary areas of research interest include wireless networks, high-performance computing, internet of things, artificial intelligence technologies - machine learning, machine translation, deep learning, data science, and data analytics. Further info on her homepage http://expert.ubd.edu.bn/sandhya.aneja. She can be contacted at sandhya.aneja@ubd.edu.bn.

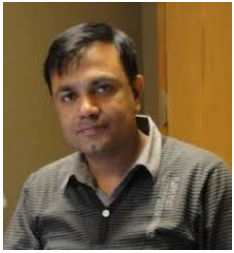 **Dr Nagender Aneja** is working as Assistant Professor at School of Digital Science, Universiti Brunei Darussalam. He did his Ph.D. in Computer Engineering from J.C. Bose University of Science and Technology YMCA, and M.E. Computer Technology and Applications from Delhi College of Engineering. He is currently working in the area of deep learning, computer vision, and natural language processing. He is also founder of ResearchID.co, further info on his homepage http://expert.ubd.edu.bn/nagender.aneja. He can be contacted at nagender.aneja@ubd.edu.bn

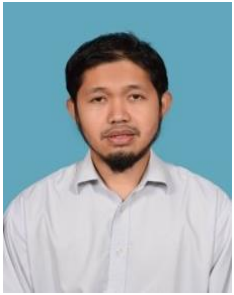 **Pg Dr Emeroylariffion Abas** received his B.Eng. Information Systems Engineering from Imperial College, London in 2001, before obtaining his Ph.D. Communication Systems in 2005 from the same institution. He is now working as an Assistant Professor in General Engineering, Faculty of Integrated Technologies, Universiti Brunei Darussalam. His present research interest are data analytic, energy systems and photonics. Further info on his homepage https://expert.ubd.edu.bn/emeroylariffion.abas. He can be contacted at emeroylariffion.abas@ubd.edu.bn.

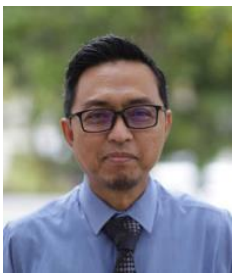 **Dr Abdul Ghani Naim** is a Senior Assistant Professor at School of Digital Science, Universiti Brunei Darussalam. He did his Ph.D. in Information Security from the Royal Holloway College, London. His present research interests include computer security, cryptography, high performance computing, and machine learning. Further info on his homepage https://expert.ubd.edu.bn/ghani.naim. He can be contacted at ghani.naim@ubd.edu.bn.